\documentclass{aa}
\usepackage{graphicx}
\usepackage{txfonts}
\newcounter{Rco}
\newcommand{\Ionst}[1]{\setcounter{Rco}{#1}\Roman{Rco}}
\newcommand{\Ion}[2]{\mbox{#1\ {\scriptsize\Ionst{#2}}}}
\newcommand{\Ionw}[3]{\mbox{#1\ {\scriptsize\Ionst{#2}}~$\lambda\,#3$\AA}}
\newcommand{\Ionwf}[3]{\mbox{[#1\,{\scriptsize\Ionst{#2}}]~$\lambda\,#3$\AA}}

\newcommand{\logg}{\mbox{$\log g$}}
\newcommand{\loggw}[1]{\mbox{$\log g\hspace{-0.5mm} =\hspace{-0.5mm}  #1$}}

\newcommand{\ab}[1]{\mbox{Fig.\,\ref{#1}}}
\newcommand{\sA}[1]{\mbox{(Fig.\,\ref{#1})}}

\newcommand{\sK}[1]{\mbox{(Sect.\,\ref{#1})}}
\newcommand{\sla}{\raisebox{-0.10em}{$\stackrel{<}{{\mbox{\tiny $\sim$}}}$}}
\newcommand{\spm}{\mbox{\raisebox{0.20em}{{\tiny \hspace{0.2mm}\mbox{$\pm$}\hspace{0.2mm}}}}}
\newcommand{\ta}[1]{\mbox{Table\,\ref{#1}}}
\newcommand{\sT}[1]{\mbox{(Table\,\ref{#1})}}
\newcommand{\Teff}{\mbox{$T_\mathrm{eff}$}}
\newcommand{\Teffw}[1]{\mbox{$\Teff\hspace{-0.5mm} =\hspace{-0.5mm} #1 \mathrm{kK}$}}

\newcommand{\ApuSS}[2]{\apss, #1, #2 }
\newcommand{\APJ}[2]{\apj, #1, #2 }

\newcommand{\ARAuA}[2]{\araa, #1, #2 }
\newcommand{\AuA}[2]{\aap, #1, #2 }

\newcommand{\MNRAS}[2]{\mnras, #1, #2 }
\newcommand{\PASP}[2]{\pasp, #1, #2 }
\begin{document}
   \title
   {A grid of synthetic ionizing spectra for very hot compact stars from NLTE model atmospheres
   }

   \author{T\@. Rauch$^{1, 2}$}
   \offprints{T\@. Rauch}
   \mail{Thomas.Rauch@sternwarte.uni-erlangen.de}
 
   \institute
    {Dr.-Remeis-Sternwarte, Sternwartstra\ss e 7, D-96049 Bamberg, Germany
    \and
     Institut f\"ur Astronomie und Astrophysik, Abteilung Astronomie, Sand 1, D-72076 T\"ubingen, Germany}
 
    \date{Received 18 December 2002 / Accepted 11 March 2003}

   \authorrunning{T\@. Rauch}
%
   \abstract{
The precise analysis of properties of planetary nebulae is strongly dependent
on good models for the stellar ionizing spectrum. Observations 
in the UV -- X-ray wavelength range
as well as NLTE
model atmosphere calculations of spectra of their exciting stars have shown
that neither blackbody fluxes nor ``standard'' NLTE atmosphere models which
are composed out of hydrogen and helium only are good approximations. Strong
differences between synthetic spectra from these compared to observed spectra
at energies higher than 54\,eV (\Ion{He}{2} ground state) can be ascribed to
the neglect of metal-line blanketing.

Realistic modeling of the emergent fluxes of hot stars in the UV -- X-ray
wavelength range requires metal-line blanketed NLTE model atmospheres which
include all elements from hydrogen up to the iron-group. For this purpose, we
present a grid (solar and halo abundance ratios) of metal-line blanketed NLTE
model atmosphere fluxes which covers the parameter range of central stars of
planetary nebulae.
            }
   \maketitle

\section{Introduction} 
\label{int}

Post-AGB stars display the hottest stage of stellar evolution just before they
enter the white dwarf cooling sequence. Effective temperatures up to 180\,kK
have been determined (e.g\@. Werner et al\@. 1997) by means of spectral analyses 
based on NLTE 
(non local thermodynamic equilibrium)
model atmosphere techniques. 

A sufficient condition for the use of LTE models in spectral analyses is,
that the collisional rates between atomic states are much larger 
than the corresponding radiative rates (cf\@. Kudritzki \& Hummer 1990). 
This is e.g\@. fulfilled in the case of ``cool'' (\Teff\,$\sla\,40\,\mathrm{kK}$) 
DA white dwarfs (Napiwotzki 1997). However, there are always NLTE effects in any 
star which are easily detectable if high-resolution high-S/N spectra, especially of  
the high energy wavelength range (UV, EUV, X-ray), are evaluated (e.g\@. Rauch 
1997 and references therein) and these effects are particularly important for
hot stars like e.g\@. central stars of planetary nebulae (CSPN). 

Observations in the UV -- X-ray wavelength range have shown that this region is
generally dominated by metal opacities. Since the flux maxima of hot stars lie 
in the EUV/X-ray wavelength \sA{bb}, and the flux in this region is crucial to  
the ionization of ambient matter, e.g\@. a surrounding planetary nebulae (PN), the employment of
realistic fluxes calculated from fully line-blanketed NLTE model atmospheres as 
ionizing spectra in photoionization models is very important. 
However, blackbody fluxes are still often in use. 

In the early 80's of the last century, the calculation of NLTE models was rudimental and
hampered by the existing insufficient numerical methods and computational capacity
(cf\@. Rauch 1997). Following Pottasch et al\@. (1978), who has shown that, based on
intermediate band photometry within 1500 - 3300\,\AA, some CSPN
have their radiation well represented by blackbodies (and others not so well), Kaler (1983) 
addressed that the major problem in PN analyses is the model selection. He invoked then
that a blackbody has at least the virtue of simplicity.

\begin{figure*}[ht]
  \resizebox{\hsize}{!}{\includegraphics{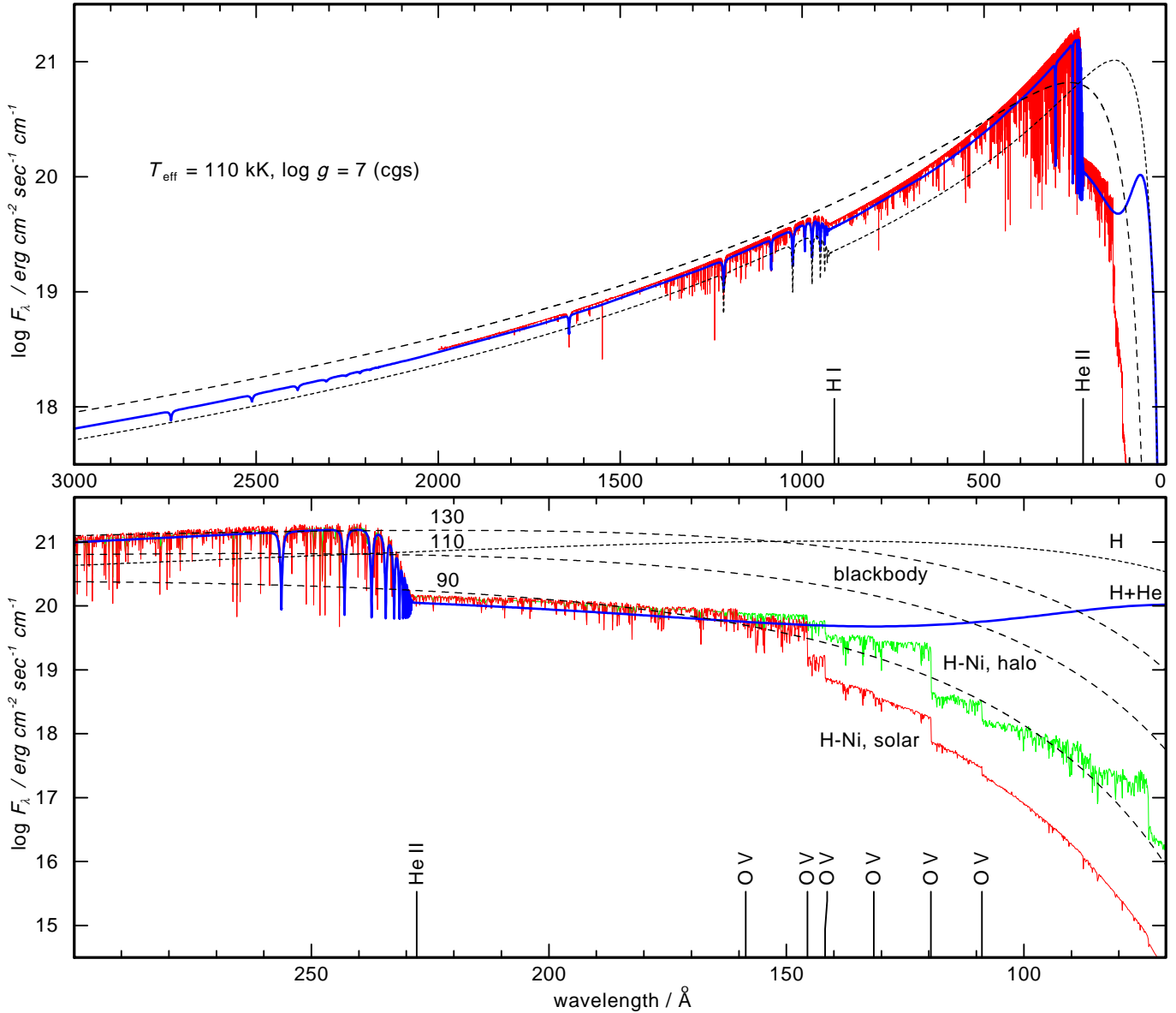}}
  \caption[]{Top: Comparison of NLTE model atmosphere fluxes 
              (\Teffw{110}, \loggw{7})
              calculated with different chemical composition.
              Short dashes: pure H model, thick line: H+He, thin line: H-Ni with
              [Z] = 0.
              Note that the main deviations of the blackbody (long dashes, 110\,kK) to the
              synthetic spectra occur shortwards of the indicated \Ion{H}{1} and \Ion{He}{2} 
              ground state thresholds.
              Lower panel: Detail of the upper panel; in addition, 
              a H-Ni model with
              [Z] = $-1$ (thin line), and three blackbody fluxes (long dashes, 90, 110, 130\,kK) are shown.
              The prominent \Ion{He}{2} and \Ion{O}{5} absorption edges are indicated. 
              Note the drastic decrease of the flux level at
              shorter wavelengths, if the opacities of metals are considered.
            }
  \label{bb}
\end{figure*}

\begin{figure}[ht]
  \resizebox{\hsize}{!}{\includegraphics{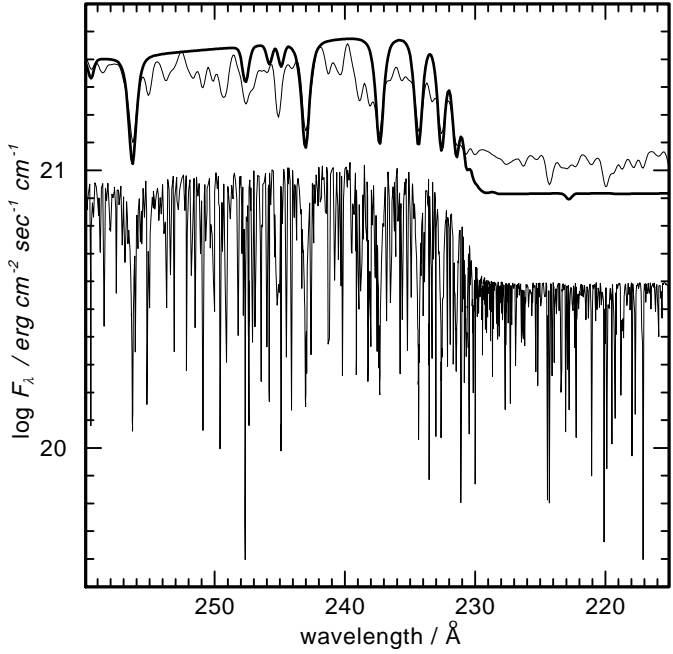}}
  \caption[]{Comparison of NLTE model atmosphere fluxes (\Teffw{130}, \loggw{7}, solar abundances) 
             without (thick line) and with consideration of iron-group opacities. For clarity, the spectra are 
             convolved with a Gaussian
             of $FWHM = 0.5\,\mathrm{\AA}$ (the original H-Ni flux is shown in addition, shifted down by 0.5 dex).
            }
  \label{irongr}
\end{figure}

\begin{figure}[ht]
  \resizebox{\hsize}{!}{\includegraphics{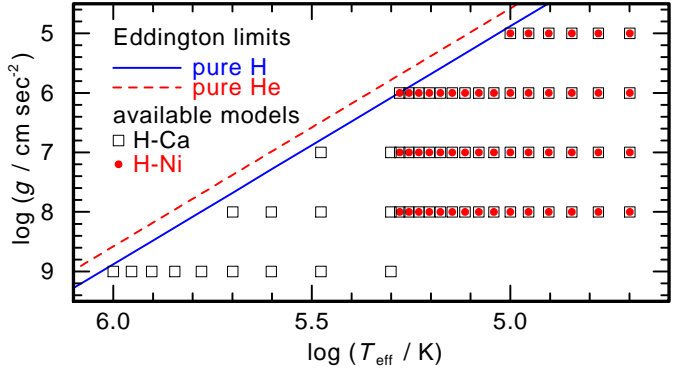}}
  \caption[]{Parameters of available NLTE model atmosphere fluxes.
             The H-Ni models are limited to \Teff\,$\leq 190\,\mathrm{kK}$ because
             Kurucz's line lists (1996) provide only data up to ionization
             stage {\sc ix}.
            }
  \label{models}
\end{figure}

\begin{table}[ht]
\caption{Statistics of model atoms used in the calculation
         of the \Teffw{110}, \loggw{7} models. NLTE is the number of levels
         treated in NLTE, RBB 
          (radiative bound-bound) 
         is the number of line transitions.
         495 additional levels are treated in LTE.} 
\label{atoms}
\begin{tabular}{lccc|clcc}
\hline
\hline
ion & NLTE & RBB &&& ion & NLTE & RBB \\
\hline
\ion{H }{i}    &  15 & 105 &&& \ion{Mg}{vi}   &   4 &   0 \\
\ion{H }{ii}   &   1 &   - &&& \ion{Mg}{vii}  &   1 &   0 \\
\ion{He}{i}    &   5 &   3 &&& \ion{Al}{iv}   &   1 &   0 \\
\ion{He}{ii}   &  16 &  84 &&& \ion{Al}{v}    &   5 &   0 \\
\ion{He}{iii}  &   1 &   - &&& \ion{Al}{vi}   &   5 &   0 \\
\ion{C }{iii}  &   3 &   1 &&& \ion{Al}{vii}  &   1 &   0 \\
\ion{C }{iv}   &  57 & 295 &&& \ion{Si}{iv}   &   1 &   0 \\
\ion{C }{v}    &   1 &   0 &&& \ion{Si}{v}    &   4 &   0 \\
\ion{N }{iv}   &   3 &   0 &&& \ion{Si}{vi}   &   2 &   0 \\
\ion{N }{v}    &  45 & 185 &&& \ion{Si}{vii}  &   1 &   0 \\
\ion{N }{vi}   &   1 &   0 &&& \ion{P }{v}    &   3 &   0 \\
\ion{O }{iv}   &   4 &   2 &&& \ion{P }{vi}   &   1 &   0 \\
\ion{O }{v}    &   5 &   1 &&& \ion{P }{vii}  &   2 &   0 \\
\ion{O }{vi}   &   1 &   0 &&& \ion{P }{viii} &   1 &   0 \\
\ion{F }{iv}   &   2 &   0 &&& \ion{S }{vi}   &   3 &   0 \\
\ion{F }{v}    &   3 &   0 &&& \ion{S }{vii}  &   6 &   0 \\
\ion{F }{vi}   &   3 &   0 &&& \ion{S }{viii} &   1 &   0 \\
\ion{F }{vii}  &   2 &   0 &&& \ion{Cl}{vi}   &   3 &   0 \\
\ion{F }{viii} &   1 &   0 &&& \ion{Cl}{vii}  &   3 &   0 \\
\ion{Ne}{iv}   &   4 &   0 &&& \ion{Cl}{viii} &   1 &   0 \\
\ion{Ne}{v}    &   9 &   3 &&& \ion{Cl}{ix}   &   1 &   0 \\
\ion{Ne}{vi}   &   6 &   5 &&& \ion{Ar}{vi}   &   4 &   0 \\
\ion{Ne}{vii}  &   4 &   2 &&& \ion{Ar}{vii}  &   3 &   0 \\
\ion{Ne}{viii} &   1 &   0 &&& \ion{Ar}{viii} &   3 &   0 \\
\ion{Na}{iv}   &   1 &   0 &&& \ion{Ar}{ix}   &   1 &   0 \\
\ion{Na}{v}    &   4 &   0 &&& \ion{K }{vi}   &   1 &   0 \\
\ion{Na}{vi}   &   5 &   0 &&& \ion{K }{vii}  &   3 &   0 \\
\ion{Na}{vii}  &   4 &   0 &&& \ion{K }{viii} &   4 &   0 \\
\ion{Na}{viii} &   1 &   0 &&& \ion{K }{ix}   &   3 &   0 \\
\ion{Mg}{iv}   &   1 &   0 &&& \ion{K }{x}    &   1 &   0 \\
\cline{5-8}
\ion{Mg}{v}    &   5 &   0 &&& total          & 286 & 676 \\
\hline 
\end{tabular} 
\end{table}

\setcounter{table}{0}
\begin{table}
\caption[]{Continued. Numbers in brackets denote individual levels and lines
           used in the statistical NLTE line-blanketing approach for the elements
           Ca - Ni.}
\begin{tabular}{rl|rr|rr}
\hline
\hline
\multicolumn{2}{c}{ion} & \multicolumn{2}{c}{NLTE} & \multicolumn{2}{c}{RBB} \\
\hline
generic & {\sc iv}   &  7 & (30\,293) & 33 & (4\,174\,474) \\
(Ca-Ni) & {\sc v}    &  7 & (20\,437) & 34 & (2\,629\,792) \\
        & {\sc vi}   &  7 & (16\,062) & 36 & (1\,763\,234) \\
        & {\sc vii}  &  7 & (12\,870) & 44 & (1\,244\,407) \\
        & {\sc viii} &  7 &  (9\,144) & 36 & (   821\,005) \\
        & {\sc ix}   &  7 & (12\,931) & 36 & (   989\,877) \\
        & {\sc  x}   &  1 &           &  0 &               \\
\hline
total   &            & 43 & (101\,737)& 219& (11\,622\,789)\\
\hline
\end{tabular}
\end{table}

\begin{table}[ht]
\caption{Example header of a flux table.}
\label{header}
{\tt
\begin{tabbing}
$\ast$ HE \= \kill
$\ast$ TUEBINGEN NLTE MODEL ATMOSPHERE PACKAGE \\
$\ast$ \\
$\ast$ DATA FILE CREATED BY O2W   AT 19.11.02 11:47:43 \\
$\ast$  MODEL CALCULATED BY PRO2  AT 20.01.01 03:59:57 \\
$\ast$ \\
$\ast$ T eff =   110000 K \\
$\ast$ log g =     7.00 (cgs) \\
$\ast$ \\
$\ast$ T eff from integrated flux:  110599 (quality check) \\
$\ast$ \\
$\ast$ NORMALIZED NUMBER FRACTIONS \\
$\ast$ \\
$\ast$ H  \> 9.079E-01 \\
$\ast$ HE \> 9.079E-02 \\
$\ast$ C  \> 3.614E-04 \\
$\ast$ N  \> 9.079E-05 \\
$\ast$ O  \> 7.211E-04 \\
$\ast$ F  \> 3.614E-08 \\
$\ast$ NE \> 3.614E-05 \\
$\ast$ NA \> 1.811E-06 \\
$\ast$ MG \> 2.871E-05 \\
$\ast$ AL \> 2.281E-06 \\
$\ast$ SI \> 3.614E-05 \\
$\ast$ P  \> 2.281E-07 \\
$\ast$ S  \> 1.439E-05 \\
$\ast$ CL \> 1.811E-07 \\
$\ast$ AR \> 4.550E-06 \\ 
$\ast$ K  \> 1.143E-07 \\
$\ast$ CA \> 4.086E-05  (GENERIC MODEL ATOM: CA - NI) \\
$\ast$ \\
$\ast$ COLUMN 1: WAVELENGTH GIVEN IN A \\
$\ast$ COLUMN 2: ASTROPHYSICAL FLUX ( F = 4H ) \\
$\ast$           F LAMBDA  GIVEN IN ERG/CM$\ast$$\ast$2/SEC/CM \\
$\ast$ ---------------------------------------------- \\
$\ast$ 19951 LINES FOLLOWING \\
\end{tabbing}
}
\end{table}

\begin{figure}[ht]
  \resizebox{\hsize}{!}{\includegraphics{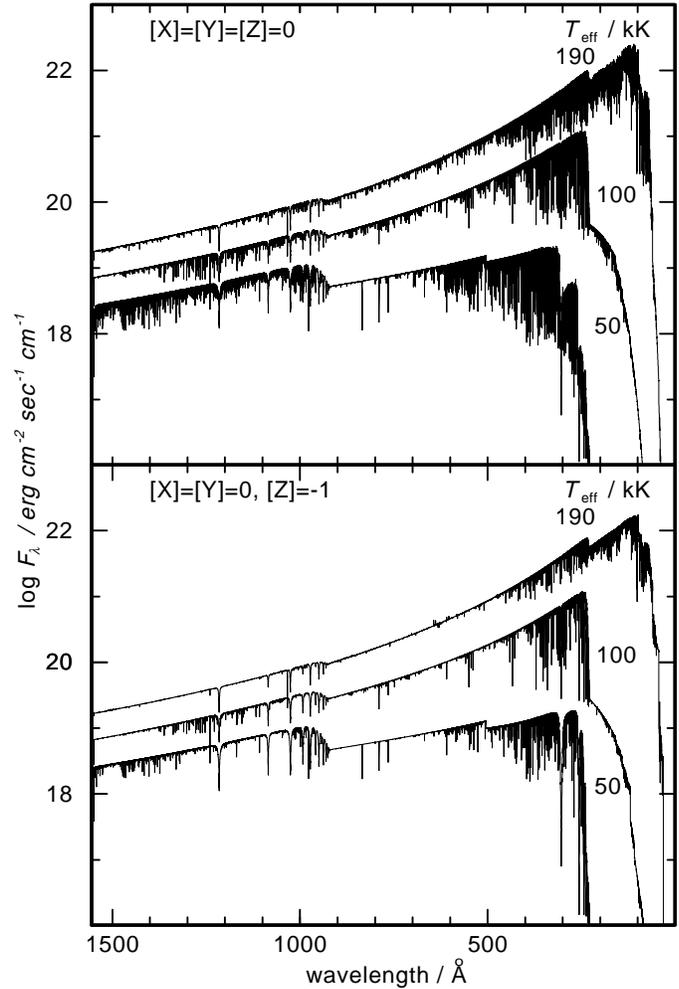}}
  \caption[]{Comparison of NLTE model atmosphere fluxes (\loggw{7}) calculated with 
             solar (top) and halo (bottom) abundance ratios at three different
             temperatures.}
  \label{svsh}
\end{figure}

The implementation of the ``accelerated lambda iteration'' (ALI) technique
in the NLTE model-atmosphere codes (Werner 1986) 
provided an efficient method to calculate realistic synthetic spectra of hot stars.
However, at that time it had been impossible to calculate extensive grids of
fully line-blanketed NLTE models due to the lack of reliable atomic data
(cf\@. Rauch 2003). Thus,
Kwok (1994) repeated Kaler's (1983) statement, that due to the variety of CSPN parameters, the
blackbody assumption is probably not worse than any other specific set of
models. In order to demonstrate the strong deviations at energies higher than 13.6\,eV and 54.4\,eV 
(\Ion{H}{1} and \Ion{He}{2} ground state thresholds, respectively), 
we show in \ab{bb} a comparison of blackbody fluxes to NLTE model atmosphere fluxes.
The weaknesses of a blackbody are obvious. Even in a H+He model with \Teffw{110}, \loggw{7}, and
a solar He/H ratio, the model flux at energies higher and lower than the \Ion{He}{2} ground state 
cannot be matched simultaneously with one blackbody flux. Instead, blackbody fluxes with
90\,kK and 130\,kK are needed, respectively. At higher energies ($\lambda < 160\,\mathrm{\AA}$), the situation gets
even worse.

NLTE model atmosphere fluxes have been successfully used, e.g\@. by Rauch et
al\@. (1994, 1996) for the construction of consistent models of 
PN and CSPN in the cases of the PNe \mbox{\object{K 1-27}} and \object{LoTr 4}. 
This approach --- consistent models of PN and CSPN --- appears to be a promising
way to improve PN analyses.

Recently, Armsdorfer et al\@. (2002) selected the spatially resolved, round 
PN \object{NGC 2438} as a test example to check for effects of CSPN models on
PN shell modeling. For their study they used the grid parameters \Teffw{110} and \loggw{7} 
which are closest to the result of Rauch et al\@. (1999, \Teffw{114\pm 10},
\loggw{6.62\pm 0.22}) from spectral analysis of the CSPN of \object{NGC 2438}. 
They found that the use of a blackbody as ionizing source leads e.g\@. to an underestimation of the
\Ionwf{O}{3}{5007} line and an overestimation of the
\Ionw{He}{2}{4686} line strength and, hence, an overestimation of the CSPN
temperature, compared to the use of a model from the H-Ca grid \sK{int}.

Rauch (1997) has investigated in detail on the impact of light metals (Li --
Ca) on NLTE model atmospheres of compact hot stars (see this paper for a more
detailed description of advances in NLTE calculations and the consideration of
metal opacities). 
He found that CNO opacities reduce the flux strongly at energies higher than
the oxygen absorption edges \sA{bb} while the opacities of light metal reduce
the flux even more, but at higher energies.
Rauch concluded that the metal opacities have a great influence
on the structure of model atmospheres. For the calculation of realistic
emergent fluxes in the EUV and X-ray range, the consideration of metals is
indispensable. Consequently, a first grid of models which are composed out of
all elements from hydrogen to calcium had been calculated with solar
([X]\footnote{[ ]: log (abundance / solar abundance)} = [Y] = [Z] = 0) and
halo ([X] = [Y] = 0, [Z] = $-1$) abundance ratios. 
This allows to interpolate or extrapolate slightly in the metal abundances [Z].
The emergent fluxes are available in tabular form at {\tt
http://astro.uni-tuebingen.de/\raisebox{1mm}{$\sim$}rauch}.

The influence of the iron-group elements was demonstrated e.g\@. by Dreizler
\& Werner (1993), Lanz \& Hubeny (1995), and Deetjen et al.\,(1999). Thus,
parallel to the calculation of the H-Ca grid, test calculations have been
performed in order to check the 
impact of the iron-group opacities (Rauch
2000, Rauch \& Deetjen 2001, Rauch 2002). 
The consideration has an influence on the atmospheric structure and thus,
also on the strengths of the absorption edges of light metals. Since significant 
changes in the flux level in the UV -- X-ray region occur \sA{irongr}, the iron-group
opacities should not be neglected. Thus, a second grid with all elements from 
hydrogen up to the iron group has been calculated. This is described in the following. 

\section{NLTE model atmospheres}
\label{nlte}

Our NLTE models (Werner 1986, 1989; Werner \& Dreizler 1999, Werner et al\@.
2003) are plane-parallel and in hydrostatic and radiative equilibrium. Hence,
our models neglect wind effect onto the spectrum, which can be significant
close to the Eddington limit (Gabler et al\@. 1989).

All elements from hydrogen to the iron-group elements can be considered
simultaneously. More than 300 levels (limited by numerical accuracy in the
used code-version) can be treated in NLTE with more than 1\,000 line
transitions calculated in detail. Millions of lines of the iron-group elements
tabulated in Kurucz (1996) and data from the Opacity Project (Seaton et al\@.
1994) are accounted for using a statistical approach and an opacity sampling
method (Deetjen 1999, {\tt
http://astro.uni-tuebingen.de/\raisebox{1mm}{$\sim$}deetjen}).

For the construction of the model atoms used in our calculations, we follow
Rauch (1997). H, He, C, N, O, and Ne are represented by small model atoms, the
other species are modeled to fit into the capacity limit of the code. As a
consequence, many of the ions of F, Na - K, \sT{atoms} are too small to
include radiative transitions, not even the resonance lines. We performed some
test calculations with larger model ions for selected atoms in order to include
at least all resonance lines of these ions. We found that these lines are very
weak, so they do not have any significant influence on the atmospheric
structure. Since our aim is not a precise spectral analysis of a given star
but the calculation of emergent fluxes as ionizing spectra, this is a good
approximation.

In contrast to Rauch (1997), Ca is not treated as a single species but
combined with the iron-group elements into a generic model atom in order to
consider all Ca lines given by Kurucz (1996). Within the generic model atom,
Ca - Ni are represented with solar abundance ratios. Depending of \Teff, the
model atoms have individually been adjusted due to the dominant ionization
stages. The statistics of representative model atoms are summarized in
\ta{atoms}. For a more detailed description of the used atomic data and the
consideration of the iron-group elements, see Rauch \& Deetjen (2003).

We calculated a grid of models within \Teffw{50 - 1\,000} and \loggw{5 - 9}
\sA{models}, with solar and halo abundances \sA{svsh}. We use
$T_{\mathrm{eff}}^{\mathrm{mod}} = \left(4\pi/\sigma \int_{0}^{\infty}
\mathcal{H}_\nu d\nu\right)^{1/4}$ for the convergence criterion
($\mathcal{H}_\nu$ is the Eddington flux). If this is closer than 2\% to the
target \Teff, the model calculation is stopped in order to stay within a
reasonable cpu time on the CRAY computers. We mention that it may take about
one day to calculate a single model.

Flux tables have subsequently been calculated within 5-2000\AA\ binned to
0.1\AA\ intervals. $T_{\mathrm{eff}}^{\mathrm{mod}}$ as well as the model's
\Teff\ and \logg\ are given in the tables' headers \sT{header}. These tables
are directly suited to be used with the photoionization code {\sc CLOUDY}
(Ferland 2000,
{\tt http://www.pa.uky.edu/\raisebox{1mm}{$\sim$}gary/cloudy/}, versions C96
Beta 4 and later). The complete flux tables, based on the frequency grids used
in the model calculations, are also available in case that a normalization to
a certain (e.g\@. V) magnitude is required.

In order to judge the necessity of a detailed consideration of iron-group
opacities in calculations of ionizing fluxes which shall be used in
photoionization models of a PN, we measured the changes of the strengths of the 
\Ion{H}{1} and \Ion{He}{2} absorption edges in the H-Ca and H-Ni fluxes. We calculated

\begin{equation}
s = \frac{
          \int^{\lambda_0}_{\lambda_0-\Delta\lambda} F_\lambda^{\mathrm{H-Ni}} \mathrm{d\lambda}
         }
         {
          \int^{\lambda_0+\Delta\lambda}_{\lambda_0} F_\lambda^{\mathrm{H-Ni}} \mathrm{d\lambda}
         }
    \left/
    \frac{
          \int^{\lambda_0}_{\lambda_0-\Delta\lambda} F_\lambda^{\mathrm{H-Ca}} \mathrm{d\lambda}
         }
         {
          \int^{\lambda_0+\Delta\lambda}_{\lambda_0} F_\lambda^{\mathrm{H-Ca}} \mathrm{d\lambda}
         }
    \right.
\end{equation}

\noindent
with $\lambda_0 = 911.5\,\mathrm{\AA}$ and $\lambda_0 = 227.8\,\mathrm{\AA}$ for
\Ion{H}{1} and \Ion{He}{2}, respectively, and $\Delta\lambda= 50\,\mathrm{\AA}$.
While the changes of the \Ion{H}{1} absorption edge are generally less than 10\%,
the strengths of the \Ion{He}{2} absorption edge may change by a factor of about
$2 - 3$ at lower temperatures. Depending on \logg, there is a maximum of $s$,
e.g\@. $s=2.3$ at \Teffw{70} and \loggw{5}. $s$ decreases towards lower and
higher \Teff\ and is about $1\spm 0.2$ at \Teff\,$> 120\,\mathrm{kK}$.  

A similar grid for hydrogen-deficient stars \sA{hdef} has been calculated.
For this grid we selected the ``typical'' abundance pattern 
(He:C:N:O=33:50:2:15 by mass)
of PG\,1159 stars (Werner 2001). This allows to estimate abundance effects onto
the emergent flux.

\begin{figure*}[ht]
  \resizebox{\hsize}{!}{\includegraphics{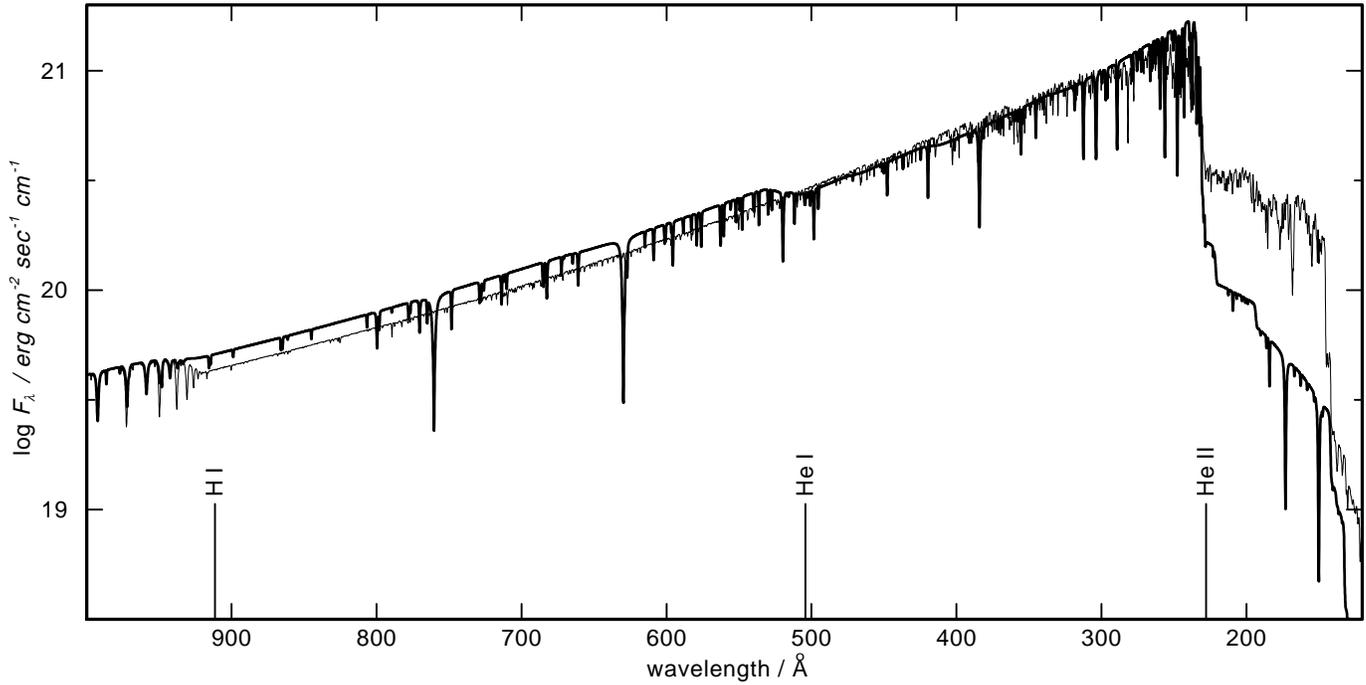}}
  \caption[]{Comparison of NLTE model atmosphere fluxes (\Teffw{110}, \loggw{6}) calculated with 
             a solar (H-Ni, thin line) and a hydrogen-deficient (He:C:N:O=33:50:2:15 by mass, thick line) composition.
             Note the strong differences at energies higher than the indicated \Ion{H}{1}, \Ion{He}{1}, and \Ion{He}{2} ground 
             state thresholds).
             For clarity, the spectra are convolved with a Gaussian of $FWHM = 0.5\,\mathrm{\AA}$.}
  \label{hdef}
\end{figure*}

\section{Conclusions}
\label{concl}

Presently available NLTE model atmosphere codes have arrived at a very high
level of sophistication (Mihalas 2003, Werner et al\@. 2003). The accelerated lambda
iteration (ALI) method represents a powerful tool to calculate realistic
metal-line blanketed atmospheres in a reasonable time, even on nowadays
available PCs 
--- these got such a tremendous number-crunching power in the 
last two years that they can now easily compete with CRAY and other
(super)computers.
The statistical approach which has been used to construct
a generic model atom for all elements from Ca to Ni \sK{nlte} has been
improved and can now be employed to construct model atoms for other light
metals in order to consider all of their lines, provided by
Kurucz (1996) and the Opacity Project (Seaton et al\@. 1994), in detail.
Thus, there is no need to reflect on blackbodies or simple model
atmospheres as ionizing sources for photoionization models. 

To summarize, the use of metal-line blanketed NLTE model atmosphere fluxes, 
which consider at least the CNO opacities in detail, as ionizing spectra for
photoionization models is highly recommended. 
The here presented fully metal-line blanketed NLTE model atmosphere fluxes 
which consider  all elements from hydrogen up to the iron group are well 
suited for this purpose.

However, the user of these model atmosphere fluxes should be aware that they
should only be used in a PN analysis if the CSPN abundances are 
in agreement with the model abundances. The use of fluxes from models
with solar abundance ratios will spoil any PN analysis if the CSPN is
hydrogen deficient because the flux shortwards of the \Ion{He}{2} ground
state edge can be different by about one dex \sA{hdef}.
A preceding spectral analysis of the CSPN which yields \Teff, \logg, and
abundance ratios and an individually
calculated small ``grid'' of NLTE model atmosphere fluxes may improve a
subsequent PN analysis --- and allows to construct consistent models
of PN and CSPN \sK{int}. 
This is also a prerequisite for reliable hydrodynamical models of the PN.

\begin{acknowledgements}
I would like to thank Klaus Werner for many helpful discussions and careful
reading of the manuscript, Jochen L\@. Deetjen for his continuous support
to calculate the iron-group opacities, the Rechenzentrum der Universit\"at Kiel, 
where the model atmosphere computations were performed on CRAY computers from 1997
to 2002, for generously awarding us the necessary computational time, 
and my colleagues from the Institut f\"ur Theoretische Physik und
Astrophysik, Abteilung Astrophysik, at Kiel University for their patience and
cooperation when the hugest batch classes on the CRAYs had been blocked for a
long time. 
This research was supported by the DLR under grants 50\,OR\,9705\,5 and
50\,OR\,0201.
\end{acknowledgements}


\begin{thebibliography}{}
\bibitem[2002]{armsd02}    Armsdorfer, B., Kimeswenger, S., \& Rauch, T\@. 
                           2002, 
                           RevMexAc 12, 180
\bibitem[1999]{deetjen99a} Deetjen, J.L\@. 
                           1999, 
                           diploma thesis, University T\"ubingen
\bibitem[1999]{deetjen99b} Deetjen, J.L., Dreizler, S., Rauch, T., \& Werner, K.
                           1999,
                           \AuA{348}{940}
\bibitem[1993]{dreizler93} Dreizler, S., \& Werner, K\@.
                           1993, 
                           \AuA{278}{199}
\bibitem[2000]{ferland00}  Ferland, G.J\@.
                           2000,
                           RevMexAc 9, 153
\bibitem[1989]{gabler89}   Gabler, R., Gabler, A., Kudritzki, R.P., Puls, J., \& Pauldrach, A.W.A\@.
                           1989,
                           \AuA{226}{162} 
\bibitem[1983]{kaler83}    Kaler, J.B\@.
                           1983,
                           \APJ{271}{188}
\bibitem[1990]{kudr1990}   Kudritzki, R.P., \& Hummer, D.G\@.
                           1990,
                           \ARAuA{28}{303}
\bibitem[1996]{kurucz96}   Kurucz, R.L\@.
                           1996,
                           in: Stellar surface structure,
                           eds\@. K.G\@. Strassmeier, J.L\@. Linsky, 
                           Kluwer, Dordrecht, p\@. 523
\bibitem[1994]{kwok94}     Kwok, S\@.
                           1994,
                           \PASP{106}{344}
\bibitem[1995]{lanz95}     Lanz, T., \& Hubeny, I\@. 
                           1995, 
                           \APJ{439}{905}
\bibitem[2003]{mihalas03}  Mihalas, D\@.
                           2003,
                           in: Workshop on Stellar Atmosphere Modeling,
                           eds\@. I\@. Hubeny, D\@. Mihalas, K\@. Werner, 
                           The ASP Conference Series Vol\@. 288 (San Francisco: ASP), p\@. 677
\bibitem[1997]{napi97}     Napiwotzki, R\@.
                           1997,
                           \AuA{322}{256}
\bibitem[1978]{pott78}     Pottasch, S.R., Wesselius, P.R., Wu, C.-C., Fieten, H., \& van Duinen, R.J\@.
                           1978,
                           \AuA{62}{95}
\bibitem[1997]{rauch97}    Rauch, T\@. 
                           1997, 
                           \AuA{320}{237}
\bibitem[2000]{rauch00}    Rauch, T\@.
                           2000,
                           \AuA{356}{665}
\bibitem[2002]{rauch02}    Rauch, T\@.
                           2002,
                           in: Proc\@. IAU Symp\@. 209. Planetary Nebulae: Their Evolution and Role in the Universe,
                           eds\@. M\@. Dopita et al.            
                           The ASP Conference Series (San Francisco: ASP), in press
\bibitem[2001]{rauch01}    Rauch, T., \& Deetjen, J.L\@.
                           2001,
                           in: Tetons 4: Galactic Structure, Stars and the Interstellar Medium,
                           eds\@. C.E\@. Woodward, M.D\@. Bicay, J.M\@. Shull,
                           The ASP Conference Series Vol\@. 231 (San Francisco: ASP), p\@. 546
\bibitem[2003]{rauch03}    Rauch T\@., \& Deetjen J.L\@.
                           2003,
                           in: Workshop on Stellar Atmosphere Modeling,
                           eds\@. I\@. Hubeny, D\@. Mihalas, K\@. Werner,
                           The ASP Conference Series Vol\@. 288 (San Francisco: ASP), p\@. 103
\bibitem[1994]{rauch94}    Rauch, T., K\"oppen, J., \& Werner, K\@.
                           1994,
                           \AuA{286}{543}
\bibitem[1996]{rauch96}    Rauch, T., K\"oppen, J., \& Werner, K\@.
                           1996,
                           \AuA{310}{613}
\bibitem[1999]{rauch99}    Rauch T., K\"oppen J., Napiwotzki R., \& Werner K\@.
                           1999,
                           \AuA{347}{169}
\bibitem[1994]{seaton94}   Seaton, M.J., Yan, Y., Mihalas, D., \& Pradhan A.K\@.
                           1994, 
                           \MNRAS{266}{805}
\bibitem[1986]{werner86}   Werner, K\@.
                           1986, 
                           \AuA{161}{177}
\bibitem[1989]{werner89}   Werner, K\@.
                           1989,
                           \AuA{226}{265}
\bibitem[2001]{werner01}   Werner, K\@.
                           2001,
                           \ApuSS{275}{27}
\bibitem[1999]{werner99}   Werner, K., \& Dreizler, S\@.
                           1999,
                           in: Computational Astrophysics,
                           eds\@. H\@. Riffert, K\@. Werner,
                           Journal of Computational and Applied Mathematics, 109, 65
\bibitem[1997]{wernerea97} Werner, K., Dreizler, S., Heber, U., et al\@.
                           1997,
                           in: Reviews in Modern Astronomy 10, 
                           ed\@. R.E\@. Schielicke, 
                           Astronomische Gesellschaft, Hamburg, p\@. 219
\bibitem[2003]{wernerea03} Werner, K., Dreizler, S, Deetjen, J.L., et al\@.
                           2003,
                           in: Workshop on Stellar Atmosphere Modeling,
                           eds\@. I\@. Hubeny, D\@. Mihalas, K\@. Werner, 
                           The ASP Conference Series Vol\@. 288 (San Francisco: ASP), p\@. 31
\end{thebibliography}
\end{document}